\documentclass{article}
\usepackage{graphicx} 

\usepackage{amsthm}
\usepackage{amsmath}
\usepackage{amssymb}
\usepackage{graphicx}
\usepackage{makecell}
\usepackage{mathtools}
\usepackage[center]{caption}
\usepackage[symbol]{footmisc}

\usepackage{perpage}
\MakePerPage{footnote}

\usepackage{setspace}
\usepackage{relsize}
\usepackage{stackengine}

\usepackage[letterpaper,margin=1in]{geometry}

\date{}

\makeatletter
\renewcommand{\fnum@figure}{\textbf{Figure \thefigure}}
\renewcommand{\fnum@table}{\textbf{Table \thetable}}
\makeatother

\usepackage{url}

\title{A functionally reversible probabilistic computing architecture enabled by interactions of current-controlled magnetic devices}
\author{
	Shreyes Nallan$^{1\ast}$,
	Jian-Gang Zhu$^{1}$\and
	\small$^{1}$Data Storage Systems Center, Carnegie Mellon University, Pittsburgh PA 15213, United States.\and
	\small$^\ast$Corresponding author. Email: shreyes@cmu.edu
}

\begin{document} 

\maketitle

\begin{abstract} \bfseries \boldmath
Probabilistic computers replace logic gates with networks of interacting random variables, creating bidirectional systems that can back-derive inputs from outputs. Such architectures enable efficient generation of random samples, implementations of novel algorithms, and natural solutions to classically hard problems such as prime factorization. We present a new physical implementation for these networks: ferromagnetic disks whose magnetization switching process is triggered by current pulses, skewed by external magnetic fields, and randomized by ambient thermal noise. We show that geometry-dependent magnetostatic interactions between these magnetic cells lead to system behavior that emulates deterministic logic gates. Furthermore, by chaining multiple ``gates," we achieve a highly accurate bidirectional one-bit full-adder, a proof of concept for complex multi-gate logic functions with reversible information flow. This analog magnetic probabilistic computer methodology improves on other implementations in speed, tunability, and energy efficiency, thereby enabling a powerful new pathway towards practical solution of classically hard problems.
\end{abstract}

In recent years, a growing body of literature has explored the use of randomness to create a novel computational framework, called “probabilistic computing" \cite{kerem-bits,borders,kerem-argument,jianping-argument,finocchio-jmmm}. A probabilistic computer is comprised of sets of interacting random variables arranged and linked to form dynamic and invertible state networks. These architectures have the potential to simulate quantum processes and run efficient and fast algorithms for integer factorization and other classically hard computing problems \cite{pcomp,kerem-is-galileo,ptechreview}. In effect, they can realize at least some of the promise of quantum computing \cite{kerem-quantum,kerem-stoquastic,kerem-vs-dwave,optics-vs-dwave} without the need for specialized hardware and extreme low-temperature environments, which are costly and inefficient \cite{nikonov,fridges}.

Such probabilistic networks have been used to provide random samples of arbitrary statistical distributions for cryptographic and machine learning applications \cite{rngapp-cryptprng,rngapp-crypto,rngapp-stoch,rngapp-spike,rngapp-ml}; to implement inference for Bayesian networks \cite{jonas,faria-bayes,kerem-bayes,senugupta-bayes}; to build so-called Ising machines to iterate solutions to classically hard problems \cite{lucas,ising-fpga,huang-sat}; and to enable reversible computing \cite{jianping,pedram-shao,borders}. In this work, we focus on the latter -- a property enabled by the fact that probabilistic computers, formed out of interactions between random bits, exhibit bidirectional information flow. Traditional computing architectures can, for instance, easily multiply two numbers together and output a result -- but they cannot start from the result and recapitulate the input numbers that led to it. This fact forms the basis of all modern cryptography. If, however, the exact same multiplication circuit was implemented with a probabilistic architecture, going backwards -- and finding the prime factors of a large number -- would be just as easy as the forwards calculation \cite{finocchio-jmmm,pedram-computer,primes}.

Most current work in the field of probabilistic computing emulates the central element of randomness with pseudorandom bit sequences generated through digital logic \cite{fpga-berkeley,fpga-janus,sutton,asic-fujitsu,asic-hitachi,asic-statica}, but this methodology is slow and extremely inefficient in device area and power utilization \cite{kerem-argument,jianping-argument,querlioz-argument}. An alternative approach is a simple, efficient, ambient-temperature source of true randomness: the interaction between thermal energy and the switching process of a magnetic cell. Even proposed architectures that do utilize magnetic switching only use magnets to generate asynchronous and unbiased random bitstreams, performing all other operations with digital logic \cite{kerem-bits,borders,kanai,jsun-fast}. Other implementations are hampered by a lack of trial-to-trial independence, probabilistic tunability, or natural physics-based interactions \cite{kent-coinflip,kent-trillions,pedram-shao,pedram-computer,jsun-strobes}. In this work, we discuss a novel approach: an entirely analog architecture for probabilistic computing, enabled by interactions between spin orbit torque magnetoresistive random access memory (SOT-MRAM) cells stabilized by in-plane uniaxial magnetocrystalline anisotropy.

\subsection*{The SOT-MRAM p-bit}
\begin{figure} 
	\centering
	\includegraphics[width=\textwidth]{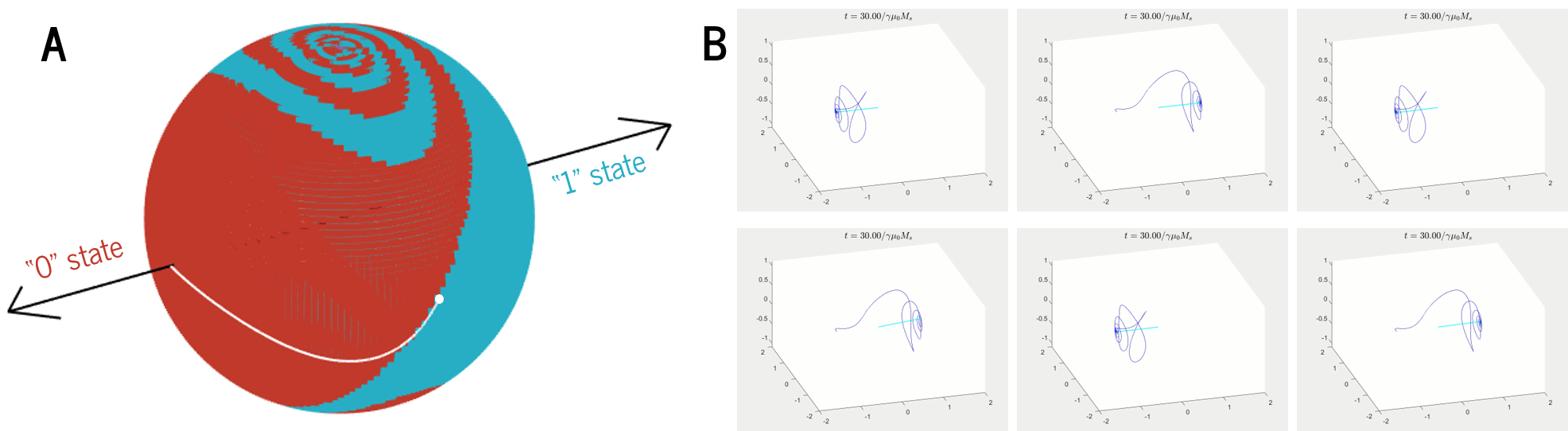}
	\caption{\textbf{Probabilistic switching dynamics in an SOT-MRAM cell.}
		(\textbf{A}) In our devices, a magnetization $\hat{m}$ starting from the red regions will damp to the left-pointing ``0" state; an $\hat{m}$ starting from the blue regions will precess to the right-pointing ``1" state. A spin pulse with magnitude $\eta$ and angle $\beta$ will send $\hat{m}$ along a preset trajectory. The path shown in white corresponds to large $\eta$ and $\beta = 90^{\circ}$; it sends $\hat{m}$ to the exact borderline between the two regions. (\textbf{B}) Time-resolved magnetization trajectories in this scenario: the final state of the SOT-MRAM cell is a random variable determined by ambient thermal fluctuations.}
	\label{fig:whirl} 
\end{figure}
Spin-orbit torque magnetic random-access memory (SOT-MRAM) is an emerging new technique within the field of spintronics \cite{liu,mram-article,three-terminal,she-review}. An SOT-MRAM device is comprised of two key layers: a heavy-metal underlayer and a ferromagnetic top layer. At their most basic level, such devices take electrical currents from the underlayer, convert them to spin currents at the heavy-metal/ferromagnet interface, and use these spin currents to write the state of the magnet. Information is stored in the direction of the magnetic moment in the top layer: we define one particular direction as the ``0" state and the diametrically opposed one as the ``1" state \cite{small-review,big-review}. In general, SOT-MRAM devices can adopt either in-plane or out-of-plane configurations \cite{liu-early,miron,fukami-xaxis,fukami-subns}.

SOT-MRAM cells are uniquely suited to meet the requirements for a probabilistic bit. They have two stable states formed by the anisotropy easy axis, which can act as the ``0” and ``1” of the probabilistic computer. This state can be regenerated on demand by an external output, the spin transfer torque arising from a current pulse. The other two required elements, randomness and cell-to-cell interaction, are provided by magnetization fluctuations from ambient thermal noise and magnetostatic stray fields emanated by the ferromagnetic disks, respectively. These latter effects are usually minor in comparison to current- and anisotropy-driven phenomena, but as explained below, the system can be manipulated such that they can play an important role.

We wish to allow in-plane stray-field-mediated interactions to be both positive (magnets pushed to the same state) and negative (magnets pushed to opposite states). To this end, we require a ferromagnetic device with stable states in $\pm\hat{x}$, within the plane of the film. The typical solution for this problem is the geometry-dependent shape anisotropy \cite{big-review,fukami-xaxis,ohno-ellipse}, but the elliptical aspect ratio required to achieve this effect imposes fundamental constraints on the areal density of our logic networks. This is particularly deleterious in our computing framework, which requires maximal freedom of geometric arrangement to function. Instead, we use SOT-MRAM devices with in-plane magnetocrystalline anisotropy, achieved through epitaxial lattice-matching at the spin Hall heterojunction; the fabrication and behavior of such devices has been covered at length in previous work \cite{me-ieee,me-conf,me-scirep}.

The switching process in this system is governed by the interplay between internal forces, encapsulated in the energetic landscape formed by anisotropy and demagnetization effective fields, and the external input, the spin torque transfer from the incoming current pulse. The energy landscape is characterized by the anisotropy energy barrier, $K$; the spin injection pulse is characterized by its magnitude $\eta$, duration $\tau$, and in-plane polarization angle $\beta$ (with respect to the easy axis). Magnetization trajectories of the top layer can broadly be divided into two segments: the current-driven phase, when the spin transfer torque dominates magnetization dynamics and drives $\hat{m}$ along a unique “switching path”; and the field-driven phase, when the effective fields take over after the end of the current pulse and damp $\hat{m}$ to a stable in-plane state.

Figure \ref{fig:whirl}A shows the regions of switching created by field-driven transient dynamics. If the magnetization points along the red regions at the time the current pulse turns off, it will eventually precess and damp back to the “0” state, corresponding to $\hat{m} = -\hat{x}$; if, instead, it points somewhere within the blue region, the effective fields will take the device to the “1” state, corresponding to $\hat{m} = +\hat{x}$.

We will apply a spin injection pulse along the in-plane hard axis – that is, with spin polarization angle $\beta = 90^{\circ}$. This will force the top-layer magnetization to rotate along the switching path shown in white in Figure \ref{fig:whirl}A. For all pulse durations beyond a certain threshold, $\hat{m}$ will be driven directly to the boundary line between the red and blue switching regions – that is, to a position exactly between the two possible equilibrium states. When the current pulse turns off, the anisotropy and demagnetization effective fields cancel each other out, enabling random thermal fluctuation – usually several orders of magnitude smaller than those two factors – to determine the final state of the magnetization. This results in a non-deterministic switching trajectory, as shown in Figure \ref{fig:whirl}B; $\hat{m}$ goes to either the “0” state or to the “1” state with equal probability. Periodic application of the spin transfer pulse, on an arbitrary clock cycle, produces a fundamentally random and independently distributed 50/50 bitstream.

\begin{figure} 
	\centering
	\includegraphics[width=\textwidth]{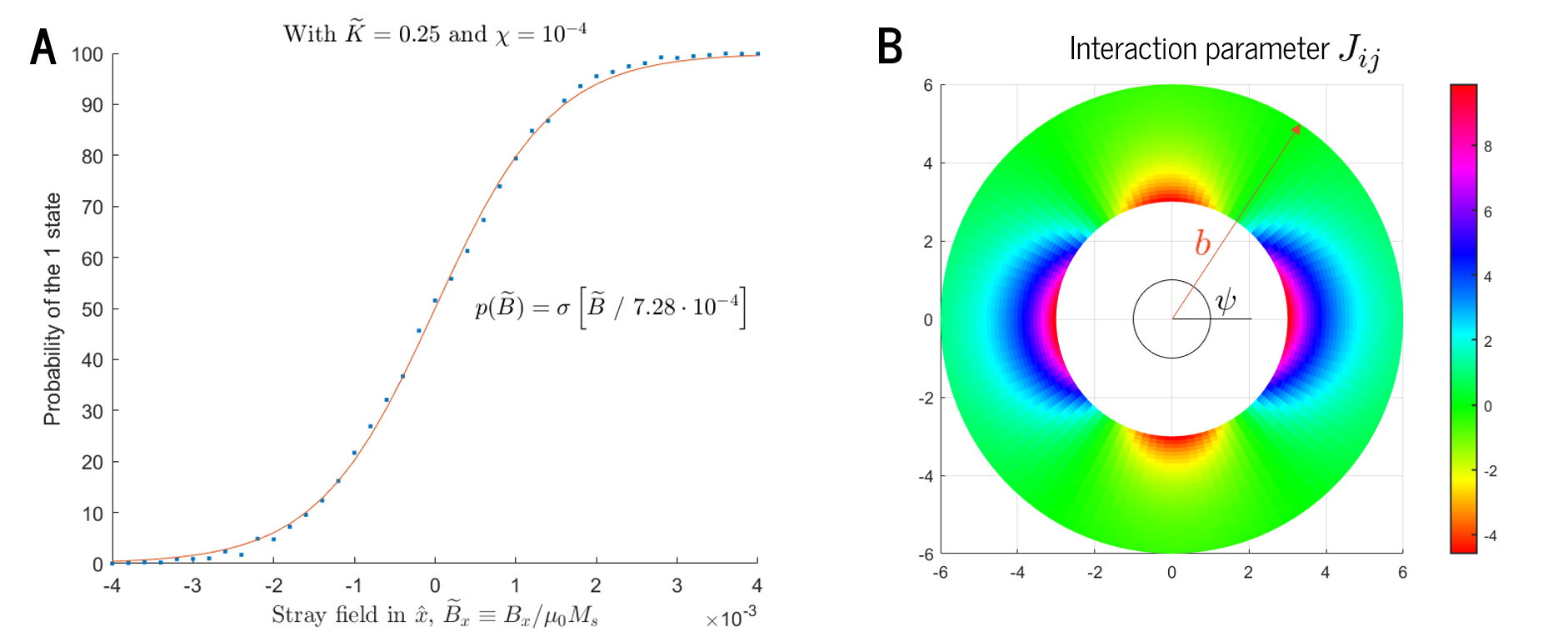}
	\caption{\textbf{Switching and interaction parameters.}
		(\textbf{A}) The probability of the magnetization falling to the ``1" state is a sigmoid with respect to the external field in the $x$-direction. The scaling factor depends on the temperature and anisotropy constant of the device. (\textbf{B}) The interaction parameter $J_{ij}$ between magnets (i.e., the input to the sigmoid) depends on the cell-to-cell distance vector, $(x_j - x_i)\hat{\mathbf{x}} + (y_j - y_i)\hat{\mathbf{y}}$, which is here represented with polar coordinates $b$ and $\psi$. Note positive $J_{ij}$ along the easy axis and negative $J_{ij}$ $90^{\circ}$ away.}
	\label{fig:probs} 
\end{figure}

Application of an external field in the easy-axis direction skews this probability, adding a lever of tunability to the system. The resulting probability characteristic, shown in Figure \ref{fig:probs}A, is sigmoidal with respect to the easy-axis field: $p = \sigma(B_x / B_0)$.

$B_0$, the sigmoidal scaling factor, depends on the following system parameters: the magnetic saturation and permeability of the ferromagnet; the relative strength of the anisotropy energy barrier; and the ratio between thermal energy and magnetic self-energy per unit volume. 

Applying a persistent spin torque along the in-plane easy axis creates a similar sigmoid-type skew in the magnetization probability, just with a different scaling factor $\eta_0$. The spin-torque bias provides a localized pathway to independently tune the probability of each SOT-MRAM device.

On the borderline between the two switching regions, anisotropy and demagnetization fields cancel each other out entirely, so even small external fields can skew the switching probability of the magnetic layer. In particular, stray fields from other, nearby magnetic cells can achieve the requisite bias – meaning that we can tie the probability of one magnet to the state of another. If the distance vector between two magnets is along the in-plane easy axis, stray field interactions will tend to keep both magnets pointing in the same direction; if, instead, the cell-to-cell distance vector is along the in-plane hard axis, the stray field will cause both magnetization states to oppose each other. As shown in Figure \ref{fig:probs}B, we can calculate a nondimensional interaction parameter between magnets $m_i$ and $m_j$ as
\begin{multline}
J_{ij} = \frac{\zeta}{4\widetilde{B}_0{((x_j - x_i)^2+(y_j - y_i)^2)}^{3/2}}\Bigg[\left(2 + \frac{3}{(x_j - x_i)^2+(y_j - y_i)^2}\right)\frac{(x_j - x_i)^2}{(x_j - x_i)^2+(y_j - y_i)^2} -\\ \left(1 + \frac{3}{4((x_j - x_i)^2+(y_j - y_i)^2)^2}\right)\frac{(y_j - y_i)^2}{(x_j - x_i)^2+(y_j - y_i)^2}\Bigg]
\label{eqn:Jij}
\end{multline}
where $(x_i,y_i)$ are the spatial coordinates of p-bit $i$.
\subsection*{Simulation setup and methodology}
The time-resolved trajectories of p-bit magnetic moments, $\hat{m_i}$, are simulated using a nondimensionalized version of the Landau-Lifshitz-Gilbert equation with Slonczewski's modification for spin transfer torque \cite{slonczewski}:
\begin{equation}
    \dfrac{d\hat{m_i}}{d\widetilde{t}} = -\dfrac{1}{1+\alpha^2}\left( \hat{m_i} \times \left[\widetilde{\mathbf{B}} + \alpha\hat{m_i} \times \widetilde{\mathbf{B}} - \alpha\widetilde{\eta}\hat{p} - \widetilde{\eta} \hat{p}\times\hat{m_i}\right]\right)
\label{eqn:LLG_nd}
\end{equation}

with $\widetilde{\mathbf{B}} \equiv \vec{B}/\mu_0M_s$, $\widetilde{\eta} = \eta / \gamma \mu_0 M_s$, and $\widetilde{t} = t \cdot \gamma\mu_0M_s$. The effective field, $\vec{B}$, contains four terms:
\begin{equation}
\vec{B} = \frac{2K}{M_s}(\hat{m} \cdot \hat{k})\hat{k} - \mu_0 M_s(\hat{m} \cdot \hat{z})\hat{z} + \vec{B}_{\mathrm{therm}} + \vec{B}_{\mathrm{ext}}\label{eqn:Beff}
\end{equation}

The first two terms are anisotropy and demagnetization effective fields, respectively. The third term is a stochastic thermal field, found through a discretization of a result from fluctuation-dissipation analysis \cite{brown,zhu-thermal} to be a Gaussian random variable with variance
\begin{equation}
    \sigma_B = \mu_0 M_s \sqrt{\dfrac{\alpha\chi}{\delta}}
\end{equation}

where $\alpha$ is the Gilbert damping, $\delta = \Delta t \cdot \gamma\mu_0 M_s$ is the simulation timestep, and $\chi \equiv (k_B T/ V) / (\frac{1}{2}\mu_0 M_s)$ is a thermal parameter characterizing the balance between thermal and magnetic energies in the system.

The fourth term is the sum of all external stray fields experienced by magnet $i$. The stray field emitted by a single ferromagnetic disk is found through a dipole superposition integral; the resulting fields from all other p-bits are added together at each timestep to find the external field at each p-bit site.

The LLG equation is numerically solved in MATLAB to find all magnetization trajectories for each p-bit cycle. Unless otherwise noted, the spin injection input is always assumed to be a rectangular pulse with magnitude $\eta$ for $0 \leq t < \tau$ and magnitude 0 otherwise.
\subsection*{Probabilistic logic gates}
\begin{figure} 
	\centering
	\includegraphics[width=\textwidth]{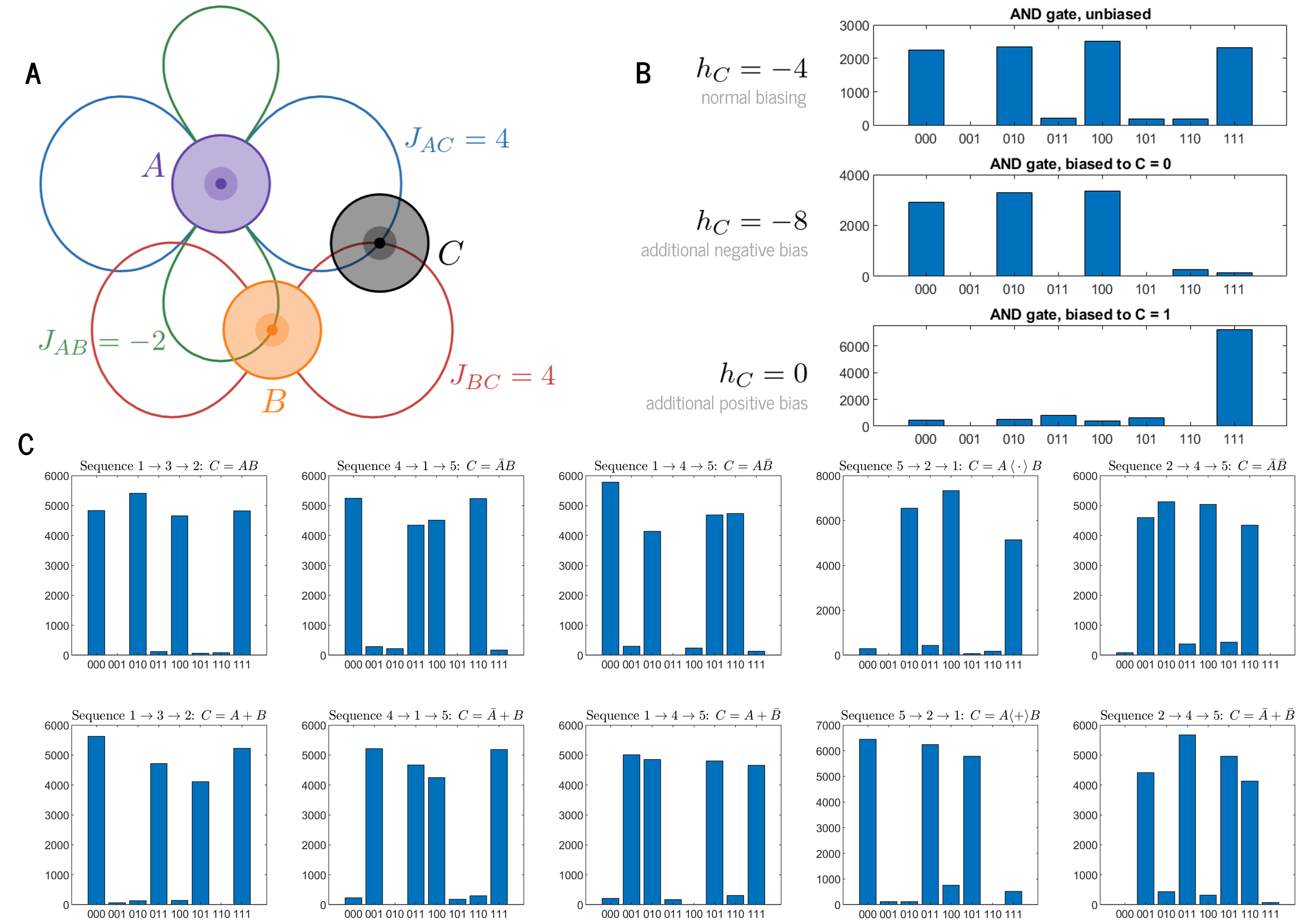}
	\caption{\textbf{Logic gate networks.}
		(\textbf{A}) The three-bit geometric arrangement corresponding to a probabilistic AND gate. (\textbf{B}) Cycling the AND gate. The system naturally falls into the $ABC$ states that match the deterministic truth table: 000, 010, 100, and 111. Fixing the output $C$ through an additional bias backpropagates to the input states $AB$: setting $C = 1$ isolates the $AB = 11$ state, while setting $C = 0$ leads the system to the other three $AB$ possibilities. (\textbf{C}) All 10 probabilistic logic functions of two inputs and one output that can be produced with the reconfigurable gate, which is a five-bit geometric arrangement. The same gate can be reconfigured on demand into an AND, OR, NAND, etc.}
	\label{fig:gates} 
\end{figure}

If a single magnet simultaneously experiences multiple cell-to-cell interactions, its probability of switching is
\begin{equation}
    p_i = \sigma\left[\sum\limits_{j \neq i}(\widetilde{\mathbf{B}}_{\mathrm{ext}}(\vec{r}_j,m_j) \cdot \hat{\mathbf{x}})\ +\ h_i\right] = \sigma\left[\sum\limits_{j \neq i}J_{ij}m_j + h_i\right]
    \label{eqn:sigsum}
\end{equation}

Where $m_j = \pm 1$ is the state of another magnet, $J_{ij}$ is calculated according to Equation \ref{eqn:Jij}, and $h_i$ is a (nondimensionalized) intrinsic spin torque bias applied to magnet $i$ during its switching cycle. Different interaction parameters -- associated with different disk arrangements -- will result in altered conditional probabilities for each individual bit and therefore altered joint probability distributions for the entire system state. In particular, we can arrange our $J_{ij}$s and $h_i$s such that our bit network matches, with arbitrarily high probabilistic accuracy, the behavior of a deterministic logic gate.

As an example, consider the magnet arrangement shown in Figure \ref{fig:gates}A. It is composed of three bits $ABC$, which are intrinsically skewed by persistent spin-torque biases $h_A = h_B = q/2$ and $h_C = -q$, and geometrically positioned to adopt the interaction parameters $J_{AB} = -q/2$ and $J_{AC} = J_{BC} = q$. These parameters produce conditional probabilities according to Equation \ref{eqn:sigsum}. When $A=1$ and $B = 1$, $C$ has a high probability of going to the “1” state (\textgreater 98\% for $q$ = 4); for all other $AB$ values, it has a low probability of “1” (\textless 2\% for $q=4$) and thus a high probability of “0”. Hence, we replicate the deterministic behavior $C=AB$. When we regenerate this bit arrangement on a clocked cycle, triggering switching in $A$, $B$, and $C$ in turn, we find that the system spends \textgreater 95\% of its time in the “correct” AND-gate states, $ABC$ = 000, 010, 100, and 111; the ``wrong" or non-AND gate states, 001, 011, 101, and 110, collectively receive \textless 5\% probability mass. Furthermore, we can add additional bias $h_C$ to fix the output state and reverse the flow of the gate: as shown in Figure \ref{fig:gates}B, if we set $C$ to 0, we force $AB$ to the three possible input states associated with this output: 00, 01, and 10. If instead we set $C$ to 1, we move all probability mass into the $AB$ = 11 state. Similar three-bit arrangements can also produce OR, NAND, and NOR gates (Figure S1, Supplementary Manuscript 1).

On first glance, stray field-mediated logic operations may seem like a finicky and unreliable methodology. However, our strategy of constructively-interfering fields, coupled with our ability to separately bias each p-bit with a $h_i$ and therefore countervail any extraneous or imbalanced magnetic fields, assists us in the development of a calibration strategy that greatly mitigates potential issues. As explained in Supplementary Manuscript 2, our logic gates are robust to device misplacement (a half-radius in any direction, and up to an entire disk radius in some directions), anisotropy constant variation of up to $\pm30\%$, and temperatures from 150 K to 600 K. In each case, the error rate of these logic gates is kept under 10\%.

An arrangement of five magnetic cells (Figure S1, Supplementary Information 1) can be reconfigured into any logic function of two inputs and one output \textit{in situ}, simply by changing the assignment of disks to variables $A$, $B$, and $C$. Depending on which set of bits we activate, we get a different set of interaction parameters $J_{ij}$ and therefore a different joint probability distribution, which can be further controlled by altering the set of intrinsic cell biases. Different combinations of these parameters result in the achievement of 10 different logic functions, including the complete set of Boolean primitives (Table S1, Supplementary Manuscript 1). The XOR and XNOR functions, which are not linearly separable, cannot be generated through this method, but we can create modified “state-avoiding” OR and AND gates (respectively), which avoid any contrary system states. All 10 joint probability distributions are shown in Figure \ref{fig:gates}C. This general-purpose reconfigurable gate can be controlled at will to emulate the behavior of any three-bit deterministic logic function.
\subsection*{Gate-to-gate communication}
The geometric gate formulation has two limitations. First, it can only create linearly separable functions of its binary inputs, precluding the XOR function, which forms the basis of many arithmetic operations. Second, it cannot provide more than about four interactions for each magnet, because the $r^{-3}$-type decay of the stray field imposes fundamental limits on the space available for cell-to-cell interaction. To create more complex logic functions under the bidirectional probabilistic framework, we must develop a way to link gate regions together and feed the output of one gate into the input of another.

One option for state transmission is an all-magnetic method of gate-to-gate communication. This can be achieved through a chain of magnetic cells, linked together through strong stray fields. A bit-flip on one end of this chain will trigger the nearest-neighbor magnet to switch in the same direction, and that magnet will trigger its nearest-neighbor, and so on until the state propagates all the way down the chain. The magnetic “pipe” can be situated along in-plane hard or easy axes, and in general, we can propagate a state in any in-plane direction by creating a stairstep pattern composed of horizontal and vertical segments. The speed of transmission is dependent on the cell-to-cell distance vector and the anisotropy barrier strength within the ferromagnets. Furthermore, we can trigger the onset of propagation and control the direction of state flow through external current-pulse inputs. This mechanism of magnetic transmission is attractive from the standpoint of energy efficiency and the implementation of an all-analog architecture, but its sensitivity to geometry, convoluted control infrastructure, and brittleness in the face of material and device variation make it somewhat impractical in its current form.

State transportation between two different probabilistic gate units does not have to be purely magnetic. Instead, this process can be more efficiently carried out through CMOS-based electronic circuits. The magnetic state of the source magnetic disk can be sensed by incorporating the ferromagnetic layer as the free layer in a magnetic tunnel junction (MTJ) structure. The reference layer of the MTJ should in this case be flux-compensated by a synthetic antiferromagnet so that it is stray field free. In this way, the magnetic state of the free layer can be sensed by CMOS circuits; some previous work has developed fast and efficient ways to accomplish this measurement \cite{cmos-zhao,cmos-shukla}. The signal generated by the MTJ, either digital or analog, can be propagated and amplified in order to drive a charge current that sets the magnetization state of the target p-bit. In this modality, the target p-bit -- the destination of the propagated signal -- may form the free layer of a spin orbit torque MTJ device. We note that a probabilistic computing architecture enabled in this way is \textit{not} equivalent to the entirely-CMOS-enabled architectures presented in previous literature; here, CMOS circuitry is only used for COPY operations between gates and not for bit-to-bit interactions and calculations within gates, greatly reducing the area, energy, and control overhead demands associated with this mechanism.

\subsection*{Large-scale bidirectional networks}
\begin{figure} 
	\centering
	\includegraphics[width=\textwidth]{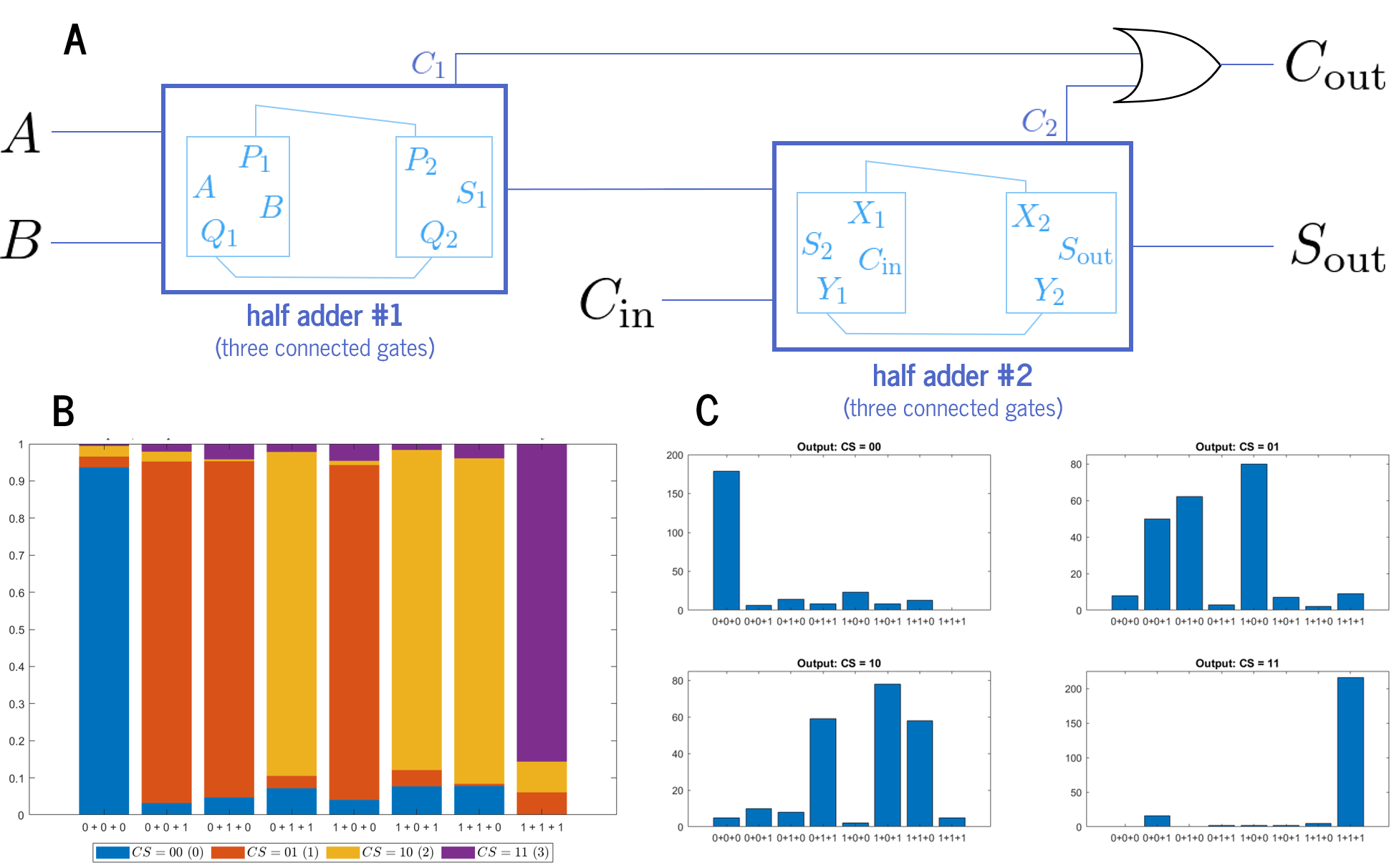}
	\caption{\textbf{A one-bit full adder.}
		(\textbf{A}) The circuit schematic, comprised of two connected half-adders and seven probabilistic gate regions. (\textbf{B}) The forward pass: we calculate the correct two-bit sum $C_{\text{out}}S_{\text{out}}$ with $\sim 90\%$ accuracy. (\textbf{C}) The backwards pass: with the same architecture, we can fix the outputs and back-derive the inputs $A$, $B$, and $C_{\text{in}}$. For instance, setting $C_{\text{out}}S_{\text{out}} = 11$ (that is, 3) leads us back to the correct input calculation: 1 + 1 + 1.}
	\label{fig:adder} 
\end{figure}

We now present a proof-of-concept device for larger-scale invertible computing. We chain together three logic operations, each with their own gate regions, to generate an XOR function: $A \oplus B = A\bar{B} + \bar{A}B$. We couple the XOR with a probabilistic AND gate to achieve a half adder, and finally chain together two half adders to create a one-bit full adder. The full probabilistic circuit schematic, featuring seven connected probabilistic gate regions, is shown in Figure \ref{fig:adder}A.

On the forward pass, we set inputs $A$, $B$, and $C_{\text{in}}$, and calculate their two-bit binary sum $C_{\text{out}} S_{\text{out}}$. We stagger our clock cycles such that we first calculate the intermediary states, then transfer these states through the communication links in the forward direction. The results of this operation are shown in Figure \ref{fig:adder}B; with high probabilistic accuracy ($\sim$90\%), we recover the correct two-bit sum output for each of the eight possible input states. Further error correction steps can increase the accuracy of this calculation, if needed, but this is beyond the scope of the work presented here.

Unlike in traditional arithmetic units, we can switch the direction of information flow \textit{in situ} -- that is, we can also set outputs and back-derive the inputs that create them. This can be done naturally with the same geometry and architecture, just with the opposite direction of transmission along the communicator units and a different sequence of bit-cycling. If we set $C_{\text{out}} S_{\text{out}} = 00$, for instance, we force almost all probability mass into the one state that creates this output, $A + B + C = 0 + 0 + 0$. Fixing $C_{\text{out}} S_{\text{out}}$ = 01 results in three possibilities, as does the output state 10; $C_{\text{out}} S_{\text{out}}$ = 11 backpropagates to the final remaining state, $A + B + C = 1 + 1 + 1$.

These results demonstrate the capability of our spintronic framework to compute non-primitive logic functions with high probabilistic accuracy in both the forward and reverse directions -- the fundamental basis of an all-magnetic probabilistic computer. We can further chain and modify the logic networks presented here to produce multi-bit adders, multipliers, and other such functional units. Eventually, we can produce a bidirectional multiplier capable of being applied to the prime factorization problem. The latter steps are, at present, beyond the scope of this work, but there is no theoretical impediment to this sort of large-scale interaction.

\subsection*{Analysis and conclusions}
We have proposed a novel framework for probabilistic computing, and we have demonstrated the operation of chained networks of SOT-MRAM devices that can implement complicated logic functions in multiple binary variables. Thanks to the cell-to-cell interactions we have developed, we can fix inputs on one side of these connected mechanisms and calculate outputs at the other; simultaneously, we can fix outputs at the other side and back-derive the inputs that create them. Notably, we we need no special processing -- no additional digital calculations or computational overhead -- for the probabilistic bits to interact with each other in the necessary manner. They do so naturally, thanks to their physical positioning and the geometric dependence of their magnetostatic stray fields. To choose different logic functions, we simply activate different SOT-MRAM cells in different cyclical sequences, and to change the direction of information flow, we simply modify the control parameters of the gate-to-gate pipes. A connected network of reconfigurable gate regions can, therefore, produce any logic function in both the forward and reverse directions. These calculations can be performed in an all-analog setting, possibly with entirely magnetic methods (if using magnetic gate-to-gate communicators).

Many current probabilistic computing implementations rely on emulated randomness through pseudorandom bit sequences, and therefore are fundamentally limited by the time, space, and energy constraints of those digital circuits. Other physical implementations, whether they use optical, resistive, or magnetic probabilistic bits, have no physical control lever to tune or interact their probabilities and therefore must farm out those calculations to separate digital-logic control circuits. The framework that we have presented here, in which the pivotal properties of randomness and interaction are “baked in” to the analog gate network, offers a promising new alternative in this field. We have demonstrated its potential for bidirectional computing, with an emphasis on the basic Boolean and arithmetic operations required for prime factorization, but we also envisage its application to the other potential use-cases of probabilistic computing, including the generation of truly random variables from arbitrary statistical distributions for cryptographic and machine-learning applications, and the implementation of novel computing algorithms in Ising machine-type frameworks.

\section*{Acknowledgments}
We thank K. Y. Camsari, A. D. Kent, P. Khalili Amiri, and P. Talatchian for helpful conversations in this field.
\paragraph*{Funding:}
All authors are funded by the Data Storage Systems Center at Carnegie Mellon University.


\newpage

\singlespacing
\pagenumbering{gobble}
\bibliographystyle{ieeetr} 
\bibliography{refs}

@article{pcomp,
    author = {Kaiser, Jan and Datta, Supriyo},
    title = {Probabilistic computing with p-bits},
    journal = {Applied Physics Letters},
    volume = {119},
    number = {15},
    pages = {150503},
    year = {2021},
    month = {10},
    issn = {0003-6951},
    doi = {10.1063/5.0067927},
    url = {https://doi.org/10.1063/5.0067927},
    eprint = {https://pubs.aip.org/aip/apl/article-pdf/doi/10.1063/5.0067927/20031145/150503_1_5.0067927.pdf},
}

@article{borders,
  title={Integer factorization using stochastic magnetic tunnel junctions},
  author={Borders, William A and Pervaiz, Ahmed Z and Fukami, Shunsuke and Camsari, Kerem Y and Ohno, Hideo and Datta, Supriyo},
  journal={Nature},
  volume={573},
  number={7774},
  pages={390--393},
  year={2019},
  publisher={Nature Publishing Group UK London}
}

@article{kerem-bits,
  title={Stochastic p-bits for invertible logic},
  author={Camsari, Kerem Yunus and Faria, Rafatul and Sutton, Brian M and Datta, Supriyo},
  journal={Physical Review X},
  volume={7},
  number={3},
  pages={031014},
  year={2017},
  publisher={APS}
}

@article{jianping,
  title={Experimental demonstration of probabilistic spin logic by magnetic tunnel junctions},
  author={Lv, Yang and Bloom, Robert P and Wang, Jian-Ping},
  journal={IEEE Magnetics Letters},
  volume={10},
  pages={1--5},
  year={2019},
  publisher={IEEE}
}

@article{kanai,
  title = {Nanosecond Random Telegraph Noise in In-Plane Magnetic Tunnel Junctions},
  author = {Hayakawa, K. and Kanai, S. and Funatsu, T. and Igarashi, J. and Jinnai, B. and Borders, W. A. and Ohno, H. and Fukami, S.},
  journal = {Phys. Rev. Lett.},
  volume = {126},
  issue = {11},
  pages = {117202},
  numpages = {6},
  year = {2021},
  month = {Mar},
  publisher = {American Physical Society},
  doi = {10.1103/PhysRevLett.126.117202},
  url = {https://link.aps.org/doi/10.1103/PhysRevLett.126.117202}
}

@article{jsun-fast,
  title={Demonstration of nanosecond operation in stochastic magnetic tunnel junctions},
  author={Safranski, Christopher and Kaiser, Jan and Trouilloud, Philip and Hashemi, Pouya and Hu, Guohan and Sun, Jonathan Z},
  journal={Nano letters},
  volume={21},
  number={5},
  pages={2040--2045},
  year={2021},
  publisher={ACS Publications}
}

@article{kent-coinflip,
  title = {Stochastic Magnetic Actuated Random Transducer Devices Based on Perpendicular Magnetic Tunnel Junctions},
  author = {Rehm, L. and Capriata, C.C.M. and Misra, S. and Smith, J.D. and Pinarbasi, M. and Malm, B.G. and Kent, A.D.},
  journal = {Phys. Rev. Appl.},
  volume = {19},
  issue = {2},
  pages = {024035},
  numpages = {9},
  year = {2023},
  month = {Feb},
  publisher = {American Physical Society},
  doi = {10.1103/PhysRevApplied.19.024035},
  url = {https://link.aps.org/doi/10.1103/PhysRevApplied.19.024035}
}

@ARTICLE{kent-trillions,
  author={Dubovski, Andre and Criss, Troy and Valli, Ahmed Sidi El and Rehm, Laura and Kent, Andrew D. and Haas, Andrew},
  journal={IEEE Magnetics Letters}, 
  title={One Trillion True Random Bits Generated With a Field-Programmable Gate Array Actuated Magnetic Tunnel Junction}, 
  year={2024},
  volume={15},
  number={},
  pages={1-4},
  doi={10.1109/LMAG.2024.3416091}}

@article{jsun-strobes,
    author = {Sun, Jonathan Z.},
    title = {Memory-compatible perpendicular magnetic tunnel junctions under bi-directional strobe write pulses: A method for generating true random number bits at high speed},
    journal = {Journal of Applied Physics},
    volume = {135},
    number = {16},
    pages = {163904},
    year = {2024},
    month = {04},
    issn = {0021-8979},
    doi = {10.1063/5.0207783},
    url = {https://doi.org/10.1063/5.0207783},
    eprint = {https://pubs.aip.org/aip/jap/article-pdf/doi/10.1063/5.0207783/19910985/163904_1_5.0207783.pdf},
}

@article{pedram-computer,
  title={An integrated-circuit-based probabilistic computer that uses voltage-controlled magnetic tunnel junctions as its entropy source},
  author={Duffee, Christian and Athas, Jordan and Shao, Yixin and Melendez, Noraica Davila and Raimondo, Eleonora and Katine, Jordan A and Camsari, Kerem Y and Finocchio, Giovanni and Khalili Amiri, Pedram},
  journal={Nature Electronics},
  pages={1--10},
  year={2025},
  publisher={Nature Publishing Group UK London}
}

@article{pedram-shao,
  title={Probabilistic computing with voltage-controlled dynamics in magnetic tunnel junctions},
  author={Shao, Yixin and Duffee, Christian and Raimondo, Eleonora and Davila, Noraica and Lopez-Dominguez, Victor and Katine, Jordan A and Finocchio, Giovanni and Amiri, Pedram Khalili},
  journal={Nanotechnology},
  volume={34},
  number={49},
  pages={495203},
  year={2023},
  publisher={IOP Publishing}
}

@ARTICLE{kerem-argument,
  author={Chowdhury, Shuvro and Grimaldi, Andrea and Aadit, Navid Anjum and Niazi, Shaila and Mohseni, Masoud and Kanai, Shun and Ohno, Hideo and Fukami, Shunsuke and Theogarajan, Luke and Finocchio, Giovanni and Datta, Supriyo and Camsari, Kerem Y.},
  journal={IEEE Journal on Exploratory Solid-State Computational Devices and Circuits}, 
  title={A Full-Stack View of Probabilistic Computing With p-Bits: Devices, Architectures, and Algorithms}, 
  year={2023},
  volume={9},
  number={1},
  pages={1-11},
  doi={10.1109/JXCDC.2023.3256981}}

@ARTICLE{jianping-argument,
  author={Zink, Brandon R. and Lv, Yang and Wang, Jian-Ping},
  journal={IEEE Journal on Exploratory Solid-State Computational Devices and Circuits}, 
  title={Review of Magnetic Tunnel Junctions for Stochastic Computing}, 
  year={2022},
  volume={8},
  number={2},
  pages={173-184},
  doi={10.1109/JXCDC.2022.3227062}}

@ARTICLE{sutton,
  author={Sutton, Brian and Faria, Rafatul and Ghantasala, Lakshmi Anirudh and Jaiswal, Risi and Camsari, Kerem Yunus and Datta, Supriyo},
  journal={IEEE Access}, 
  title={Autonomous Probabilistic Coprocessing With Petaflips per Second}, 
  year={2020},
  volume={8},
  number={},
  pages={157238-157252},
  doi={10.1109/ACCESS.2020.3018682}}

@article{jonas,
  title={Building fast bayesian computing machines out of intentionally stochastic, digital parts},
  author={Mansinghka, Vikash and Jonas, Eric},
  journal={arXiv preprint arXiv:1402.4914},
  year={2014}
}

@article{ptechreview,
  title={Ising machines as hardware solvers of combinatorial optimization problems},
  author={Mohseni, Naeimeh and McMahon, Peter L and Byrnes, Tim},
  journal={Nature Reviews Physics},
  volume={4},
  number={6},
  pages={363--379},
  year={2022},
  publisher={Nature Publishing Group UK London}
}

@article{fpga-janus,
  title={Janus II: A new generation application-driven computer for spin-system simulations},
  author={Baity-Jesi, Marco and Ba{\~n}os, Rachel A and Cruz, Andres and Fernandez, Luis Antonio and Gil-Narvi{\'o}n, Jos{\'e} Miguel and Gordillo-Guerrero, Antonio and Iniguez, David and Maiorano, Andrea and Mantovani, Filippo and Marinari, Enzo and others},
  journal={Computer Physics Communications},
  volume={185},
  number={2},
  pages={550--559},
  year={2014},
  publisher={Elsevier}
}

@inproceedings{asic-hitachi,
  title={2.6 A 2$\times$ 30k-spin multichip scalable annealing processor based on a processing-in-memory approach for solving large-scale combinatorial optimization problems},
  author={Takemoto, Takashi and Hayashi, Masato and Yoshimura, Chihiro and Yamaoka, Masanao},
  booktitle={2019 IEEE International Solid-State Circuits Conference-(ISSCC)},
  pages={52--54},
  year={2019},
  organization={IEEE}
}

@ARTICLE{asic-fujitsu,
  
AUTHOR={Aramon, Maliheh  and Rosenberg, Gili  and Valiante, Elisabetta  and Miyazawa, Toshiyuki  and Tamura, Hirotaka  and Katzgraber, Helmut G. },
         
TITLE={Physics-Inspired Optimization for Quadratic Unconstrained Problems Using a Digital Annealer},
        
JOURNAL={Frontiers in Physics},
        
VOLUME={Volume 7 - 2019},

YEAR={2019},

URL={https://www.frontiersin.org/journals/physics/articles/10.3389/fphy.2019.00048},

DOI={10.3389/fphy.2019.00048},

ISSN={2296-424X}}

@INPROCEEDINGS{asic-statica,
  author={Yamamoto, Kasho and Ando, Kota and Mertig, Normann and Takemoto, Takashi and Yamaoka, Masanao and Teramoto, Hiroshi and Sakai, Akira and Takamaeda-Yamazaki, Shinya and Motomura, Masato},
  booktitle={2020 IEEE International Solid-State Circuits Conference - (ISSCC)}, 
  title={7.3 STATICA: A 512-Spin 0.25M-Weight Full-Digital Annealing Processor with a Near-Memory All-Spin-Updates-at-Once Architecture for Combinatorial Optimization with Complete Spin-Spin Interactions}, 
  year={2020},
  volume={},
  number={},
  pages={138-140},
  doi={10.1109/ISSCC19947.2020.9062965}}

@article{fpga-berkeley,
  title={Ising model optimization problems on a FPGA accelerated restricted Boltzmann machine},
  author={Patel, Saavan and Chen, Lili and Canoza, Philip and Salahuddin, Sayeef},
  journal={arXiv preprint arXiv:2008.04436},
  year={2020}
}

@article{finocchio-jmmm,
title = {The promise of spintronics for unconventional computing},
journal = {Journal of Magnetism and Magnetic Materials},
volume = {521},
pages = {167506},
year = {2021},
issn = {0304-8853},
doi = {https://doi.org/10.1016/j.jmmm.2020.167506},
url = {https://www.sciencedirect.com/science/article/pii/S0304885320324732},
author = {Giovanni Finocchio and Massimiliano {Di Ventra} and Kerem Y. Camsari and Karin Everschor-Sitte and Pedram {Khalili Amiri} and Zhongming Zeng}
}

@article{querlioz-argument,
  title = {Low-Energy Truly Random Number Generation with Superparamagnetic Tunnel Junctions for Unconventional Computing},
  author = {Vodenicarevic, D. and Locatelli, N. and Mizrahi, A. and Friedman, J. S. and Vincent, A. F. and Romera, M. and Fukushima, A. and Yakushiji, K. and Kubota, H. and Yuasa, S. and Tiwari, S. and Grollier, J. and Querlioz, D.},
  journal = {Phys. Rev. Appl.},
  volume = {8},
  issue = {5},
  pages = {054045},
  numpages = {9},
  year = {2017},
  month = {Nov},
  publisher = {American Physical Society},
  doi = {10.1103/PhysRevApplied.8.054045},
  url = {https://link.aps.org/doi/10.1103/PhysRevApplied.8.054045}
}

@article{rngapp-cryptprng,
  title={Pseudorandom numbers},
  author={Lagarias, Jeffrey C},
  journal={Statistical Science},
  volume={8},
  number={1},
  pages={31--39},
  year={1993},
  publisher={Institute of Mathematical Statistics}
}

@Inbook{rngapp-crypto,
author="Stip{\v{c}}evi{\'{c}}, Mario
and Ko{\c{c}}, {\c{C}}etin Kaya",
title="True Random Number Generators",
bookTitle="Open Problems in Mathematics and Computational Science",
year="2014",
publisher="Springer International Publishing",
address="Cham",
pages="275--315",
isbn="978-3-319-10683-0",
doi="10.1007/978-3-319-10683-0_12",
url="https://doi.org/10.1007/978-3-319-10683-0_12"
}

@ARTICLE{rngapp-stoch,
  author={Hamilton, Tara Julia and Afshar, Saeed and van Schaik, André and Tapson, Jonathan},
  journal={Proceedings of the IEEE}, 
  title={Stochastic Electronics: A Neuro-Inspired Design Paradigm for Integrated Circuits}, 
  year={2014},
  volume={102},
  number={5},
  pages={843-859},
  doi={10.1109/JPROC.2014.2310713}}

@ARTICLE{rngapp-spike,
  author={Maass, Wolfgang},
  journal={Proceedings of the IEEE}, 
  title={Noise as a Resource for Computation and Learning in Networks of Spiking Neurons}, 
  year={2014},
  volume={102},
  number={5},
  pages={860-880},
  doi={10.1109/JPROC.2014.2310593}}

@ARTICLE{rngapp-ml,
  
AUTHOR={Fu, Zhenxiao  and Tang, Yi  and Zhao, Xi  and Lu, Kai  and Dong, Yemin  and Shukla, Amit  and Zhu, Zhifeng  and Yang, Yumeng },
         
TITLE={An Overview of Spintronic True Random Number Generator},
        
JOURNAL={Frontiers in Physics},
        
VOLUME={Volume 9 - 2021},

YEAR={2021},

URL={https://www.frontiersin.org/journals/physics/articles/10.3389/fphy.2021.638207},

DOI={10.3389/fphy.2021.638207},

ISSN={2296-424X}}

@ARTICLE{faria-bayes,
  
AUTHOR={Faria, Rafatul  and Kaiser, Jan  and Camsari, Kerem Y.  and Datta, Supriyo },
         
TITLE={Hardware Design for Autonomous Bayesian Networks},
        
JOURNAL={Frontiers in Computational Neuroscience},
        
VOLUME={Volume 15 - 2021},

YEAR={2021},

URL={https://www.frontiersin.org/journals/computational-neuroscience/articles/10.3389/fncom.2021.584797},

DOI={10.3389/fncom.2021.584797},

ISSN={1662-5188}}

@article{senugupta-bayes,
  title={Stochastic spin-orbit torque devices as elements for bayesian inference},
  author={Shim, Yong and Chen, Shuhan and Sengupta, Abhronil and Roy, Kaushik},
  journal={Scientific reports},
  volume={7},
  number={1},
  pages={14101},
  year={2017},
  publisher={Nature Publishing Group UK London}
}

@article{kerem-bayes,
    author = {Faria, Rafatul and Camsari, Kerem Y. and Datta, Supriyo},
    title = {Implementing Bayesian networks with embedded stochastic MRAM},
    journal = {AIP Advances},
    volume = {8},
    number = {4},
    pages = {045101},
    year = {2018},
    month = {04},
    issn = {2158-3226},
    doi = {10.1063/1.5021332},
    url = {https://doi.org/10.1063/1.5021332},
    eprint = {https://pubs.aip.org/aip/adv/article-pdf/doi/10.1063/1.5021332/19733984/045101_1_online.pdf},
}

@ARTICLE{lucas,
  
AUTHOR={Lucas, Andrew },
         
TITLE={Ising formulations of many NP problems},
        
JOURNAL={Frontiers in Physics},
        
VOLUME={Volume 2 - 2014},

YEAR={2014},

URL={https://www.frontiersin.org/journals/physics/articles/10.3389/fphy.2014.00005},

DOI={10.3389/fphy.2014.00005},

ISSN={2296-424X}}

@article{ising-fpga,
  title={Ising model optimization problems on a FPGA accelerated restricted Boltzmann machine},
  author={Patel, Saavan and Chen, Lili and Canoza, Philip and Salahuddin, Sayeef},
  journal={arXiv preprint arXiv:2008.04436},
  year={2020}
}

@article{huang-sat,
  title={Augmenting an electronic Ising machine to effectively solve boolean satisfiability},
  author={Sharma, Anshujit and Burns, Matthew and Hahn, Andrew and Huang, Michael},
  journal={Scientific Reports},
  volume={13},
  number={1},
  pages={22858},
  year={2023},
  publisher={Nature Publishing Group UK London}
}

@article{primes,
    author = {Traversa, Fabio L. and Di Ventra, Massimiliano},
    title = {Polynomial-time solution of prime factorization and NP-complete problems with digital memcomputing machines},
    journal = {Chaos: An Interdisciplinary Journal of Nonlinear Science},
    volume = {27},
    number = {2},
    pages = {023107},
    year = {2017},
    month = {02},
    issn = {1054-1500},
    doi = {10.1063/1.4975761},
    url = {https://doi.org/10.1063/1.4975761},
    eprint = {https://pubs.aip.org/aip/cha/article-pdf/doi/10.1063/1.4975761/19777635/023107_1_online.pdf},
}

@ARTICLE{kerem-is-galileo,
  author={Camsari, Kerem and Datta, Supriyo},
  journal={IEEE Spectrum}, 
  title={Dialogue Concerning the Two Chief Computing Systems: Imagine yourself on a flight talking to an engineer about a scheme that straddles classical and quantum}, 
  year={2021},
  volume={58},
  number={4},
  pages={30-35},
  doi={10.1109/MSPEC.2021.9393992}}

@INPROCEEDINGS{kerem-quantum,
  author={Chowdhury, S. and Datta, S. and Camsari, K. Y.},
  booktitle={2019 IEEE International Electron Devices Meeting (IEDM)}, 
  title={A Probabilistic Approach to Quantum Inspired Algorithms}, 
  year={2019},
  volume={},
  number={},
  pages={37.5.1-37.5.4},
  doi={10.1109/IEDM19573.2019.8993655}}

@article{nikonov,
  title={Stochastic magnetic circuits rival quantum computing},
  author={Nikonov, Dmitri E},
  journal={Nature},
  year={2019},
  volume={573},
  pages={351--352},
  publisher={Nature Publishing Group UK London}
}

@article{kerem-stoquastic,
  title = {Scalable Emulation of Sign-Problem--Free Hamiltonians with Room-Temperature $p$-bits},
  author = {Camsari, Kerem Y. and Chowdhury, Shuvro and Datta, Supriyo},
  journal = {Phys. Rev. Appl.},
  volume = {12},
  issue = {3},
  pages = {034061},
  numpages = {14},
  year = {2019},
  month = {Sep},
  publisher = {American Physical Society},
  doi = {10.1103/PhysRevApplied.12.034061},
  url = {https://link.aps.org/doi/10.1103/PhysRevApplied.12.034061}
}

@article{
optics-vs-dwave,
author = {Ryan Hamerly  and Takahiro Inagaki  and Peter L. McMahon  and Davide Venturelli  and Alireza Marandi  and Tatsuhiro Onodera  and Edwin Ng  and Carsten Langrock  and Kensuke Inaba  and Toshimori Honjo  and Koji Enbutsu  and Takeshi Umeki  and Ryoichi Kasahara  and Shoko Utsunomiya  and Satoshi Kako  and Ken-ichi Kawarabayashi  and Robert L. Byer  and Martin M. Fejer  and Hideo Mabuchi  and Dirk Englund  and Eleanor Rieffel  and Hiroki Takesue  and Yoshihisa Yamamoto },
title = {Experimental investigation of performance differences between coherent Ising machines and a quantum annealer},
journal = {Science Advances},
volume = {5},
number = {5},
pages = {eaau0823},
year = {2019},
doi = {10.1126/sciadv.aau0823},
URL = {https://www.science.org/doi/abs/10.1126/sciadv.aau0823},
eprint = {https://www.science.org/doi/pdf/10.1126/sciadv.aau0823}}

@article{kerem-vs-dwave,
  title={Accelerated quantum Monte Carlo with probabilistic computers},
  author={Chowdhury, Shuvro and Camsari, Kerem Y and Datta, Supriyo},
  journal={Communications Physics},
  volume={6},
  number={1},
  pages={85},
  year={2023},
  publisher={Nature Publishing Group UK London}
}

@ARTICLE{fridges,
  author={Holmes, D. Scott and Ripple, Andrew L. and Manheimer, Marc A.},
  journal={IEEE Transactions on Applied Superconductivity}, 
  title={Energy-Efficient Superconducting Computing—Power Budgets and Requirements}, 
  year={2013},
  volume={23},
  number={3},
  pages={1701610-1701610},
  doi={10.1109/TASC.2013.2244634}}

@article{slonczewski,
title = {Current-driven excitation of magnetic multilayers},
journal = {Journal of Magnetism and Magnetic Materials},
volume = {159},
number = {1},
pages = {L1-L7},
year = {1996},
issn = {0304-8853},
doi = {https://doi.org/10.1016/0304-8853(96)00062-5},
url = {https://www.sciencedirect.com/science/article/pii/0304885396000625},
author = {J.C. Slonczewski}
}

@article{mram-article,
  title={MRAM gets closer to the core},
  author={Liu, Yizhou and Yu, Guoqiang},
  journal={Nature Electronics},
  volume={2},
  number={12},
  pages={555--556},
  year={2019},
  publisher={Nature Publishing Group UK London}
}

@ARTICLE{three-terminal,
  author={Lee, Seo-Won and Lee, Kyung-Jin},
  journal={Proceedings of the IEEE}, 
  title={Emerging Three-Terminal Magnetic Memory Devices}, 
  year={2016},
  volume={104},
  number={10},
  pages={1831-1843},
  doi={10.1109/JPROC.2016.2543782}}

@article{liu,
author = {Luqiao Liu  and Chi-Feng Pai  and Y. Li  and H. W. Tseng  and D. C. Ralph  and R. A. Buhrman },
title = {Spin-Torque Switching with the Giant Spin Hall Effect of Tantalum},
journal = {Science},
volume = {336},
number = {6081},
pages = {555-558},
year = {2012},
doi = {10.1126/science.1218197},
URL = {https://www.science.org/doi/abs/10.1126/science.1218197},
eprint = {https://www.science.org/doi/pdf/10.1126/science.1218197}}

@article{she-review,
  title = {Spin Hall effects},
  author = {Sinova, Jairo and Valenzuela, Sergio O. and Wunderlich, J. and Back, C. H. and Jungwirth, T.},
  journal = {Rev. Mod. Phys.},
  volume = {87},
  issue = {4},
  pages = {1213--1260},
  numpages = {47},
  year = {2015},
  month = {Oct},
  publisher = {American Physical Society},
  doi = {10.1103/RevModPhys.87.1213},
  url = {https://link.aps.org/doi/10.1103/RevModPhys.87.1213}
}

@ARTICLE{big-review,
  author={Shao, Qiming and Li, Peng and Liu, Luqiao and Yang, Hyunsoo and Fukami, Shunsuke and Razavi, Armin and Wu, Hao and Wang, Kang and Freimuth, Frank and Mokrousov, Yuriy and Stiles, Mark D. and Emori, Satoru and Hoffmann, Axel and Åkerman, Johan and Roy, Kaushik and Wang, Jian-Ping and Yang, See-Hun and Garello, Kevin and Zhang, Wei},
  journal={IEEE Transactions on Magnetics}, 
  title={Roadmap of Spin–Orbit Torques}, 
  year={2021},
  volume={57},
  number={7},
  pages={1-39},
  doi={10.1109/TMAG.2021.3078583}}

@INPROCEEDINGS{fukami-subns,
  author={Fukami, Shunsuke and Anekawa, Tetsuro and Ohkawara, Ayato and Chaoliang Zhang and Ohno, Hideo},
  booktitle={2016 IEEE Symposium on VLSI Technology}, 
  title={A sub-ns three-terminal spin-orbit torque induced switching device}, 
  year={2016},
  volume={},
  number={},
  pages={1-2},
  doi={10.1109/VLSIT.2016.7573379}}

@article{fukami-xaxis,
  title={A spin--orbit torque switching scheme with collinear magnetic easy axis and current configuration},
  author={Fukami, Shunsuke and Anekawa, T and Zhang, C and Ohno, H},
  journal={nature nanotechnology},
  volume={11},
  number={7},
  pages={621--625},
  year={2016},
  publisher={Nature Publishing Group UK London}
}

@article{liu-early,
  title = {Current-Induced Switching of Perpendicularly Magnetized Magnetic Layers Using Spin Torque from the Spin Hall Effect},
  author = {Liu, Luqiao and Lee, O. J. and Gudmundsen, T. J. and Ralph, D. C. and Buhrman, R. A.},
  journal = {Phys. Rev. Lett.},
  volume = {109},
  issue = {9},
  pages = {096602},
  numpages = {5},
  year = {2012},
  month = {Aug},
  publisher = {American Physical Society},
  doi = {10.1103/PhysRevLett.109.096602},
  url = {https://link.aps.org/doi/10.1103/PhysRevLett.109.096602}
}

@article{miron,
  title={Perpendicular switching of a single ferromagnetic layer induced by in-plane current injection},
  author={Miron, Ioan Mihai and Garello, Kevin and Gaudin, Gilles and Zermatten, Pierre-Jean and Costache, Marius V and Auffret, St{\'e}phane and Bandiera, S{\'e}bastien and Rodmacq, Bernard and Schuhl, Alain and Gambardella, Pietro},
  journal={Nature},
  volume={476},
  number={7359},
  pages={189--193},
  year={2011},
  publisher={Nature Publishing Group UK London}
}

@article{ohno-ellipse,
    author = {Takahashi, Yu and Takeuchi, Yutaro and Zhang, Chaoliang and Jinnai, Butsurin and Fukami, Shunsuke and Ohno, Hideo},
    title = {Spin-orbit torque-induced switching of in-plane magnetized elliptic nanodot arrays with various easy-axis directions measured by differential planar Hall resistance},
    journal = {Applied Physics Letters},
    volume = {114},
    number = {1},
    pages = {012410},
    year = {2019},
    month = {01},
    issn = {0003-6951},
    doi = {10.1063/1.5075542},
    url = {https://doi.org/10.1063/1.5075542},
    eprint = {https://pubs.aip.org/aip/apl/article-pdf/doi/10.1063/1.5075542/13015016/012410_1_online.pdf},
}

@INPROCEEDINGS{me-conf,
  author={Nallan, Shreyes and Zhu, Jian-Gang},
  booktitle={2024 IEEE International Magnetic Conference - Short papers (INTERMAG Short papers)}, 
  title={The Effect of Thermal Fields on Spin Hall Switching in Devices Stabilized by In-Plane Magnetocrystalline Anisotropy}, 
  year={2024},
  volume={},
  number={},
  pages={1-2},
  doi={10.1109/INTERMAGShortPapers61879.2024.10576981}}

@ARTICLE{me-ieee,
  author={Nallan, Shreyes and Zhu, Jian-Gang},
  journal={IEEE Transactions on Magnetics}, 
  title={Spin Hall Switching Enabled by Uniaxial In-Plane Magnetocrystalline Anisotropy}, 
  year={2023},
  volume={59},
  number={11},
  pages={1-5},
  doi={10.1109/TMAG.2023.3286383}}

@article{me-scirep,
  title={Manipulating transient SOT-MRAM switching dynamics for efficiency improvement and probabilistic switching},
  author={Nallan, Shreyes and Zhu, Jian-Gang},
  journal={Scientific Reports},
  volume={15},
  number={1},
  pages={38182},
  year={2025},
  publisher={Nature Publishing Group UK London}
}

@article{zhu-thermal,
    author = {Zhu, Jian-Gang},
    title = {Thermal magnetic noise and spectra in spin valve heads},
    journal = {Journal of Applied Physics},
    volume = {91},
    number = {10},
    pages = {7273-7275},
    year = {2002},
    month = {05},
    issn = {0021-8979},
    doi = {10.1063/1.1452675},
    url = {https://doi.org/10.1063/1.1452675},
    eprint = {https://pubs.aip.org/aip/jap/article-pdf/91/10/7273/19066486/7273_1_online.pdf},
}

@article{brown,
  title = {Thermal Fluctuations of a Single-Domain Particle},
  author = {Brown, William Fuller},
  journal = {Phys. Rev.},
  volume = {130},
  issue = {5},
  pages = {1677--1686},
  numpages = {0},
  year = {1963},
  month = {Jun},
  publisher = {American Physical Society},
  doi = {10.1103/PhysRev.130.1677},
  url = {https://link.aps.org/doi/10.1103/PhysRev.130.1677}
}

@article{small-review,
author = {Apalkov, Dmytro and Khvalkovskiy, Alexey and Watts, Steven and Nikitin, Vladimir and Tang, Xueti and Lottis, Daniel and Moon, Kiseok and Luo, Xiao and Chen, Eugene and Ong, Adrian and Driskill-Smith, Alexander and Krounbi, Mohamad},
title = {Spin-transfer torque magnetic random access memory (STT-MRAM)},
year = {2013},
issue_date = {May 2013},
publisher = {Association for Computing Machinery},
address = {New York, NY, USA},
volume = {9},
number = {2},
issn = {1550-4832},
url = {https://doi.org/10.1145/2463585.2463589},
doi = {10.1145/2463585.2463589},
journal = {J. Emerg. Technol. Comput. Syst.},
month = may,
articleno = {13},
numpages = {35}
}

@ARTICLE{cmos-zhao,
  author={Zhao, Weisheng and Chappert, Claude and Javerliac, Virgile and Noziere, Jean-Pierre},
  journal={IEEE Transactions on Magnetics}, 
  title={High Speed, High Stability and Low Power Sensing Amplifier for MTJ/CMOS Hybrid Logic Circuits}, 
  year={2009},
  volume={45},
  number={10},
  pages={3784-3787},
  doi={10.1109/TMAG.2009.2024325}}

@ARTICLE{cmos-shukla,
  author={Shukla, Alok Kumar and Dhull, Seema and Nisar, Arshid and Soni, Sandeep and Bindal, Namita and Kaushik, Brajesh Kumar},
  journal={IEEE Open Journal of Nanotechnology}, 
  title={Novel Radiation Hardened SOT-MRAM Read Circuit for Multi-Node Upset Tolerance}, 
  year={2022},
  volume={3},
  number={},
  pages={78-84},
  doi={10.1109/OJNANO.2022.3181040}}
\end{document}